\documentclass{article}

\usepackage{PRIMEarxiv}

\usepackage[utf8]{inputenc} % allow utf-8 input
\usepackage[T1]{fontenc}    % use 8-bit T1 fonts
\usepackage{hyperref}       % hyperlinks
\usepackage{url}            % simple URL typesetting
\usepackage{booktabs}       % professional-quality tables
\usepackage{amsfonts}       % blackboard math symbols
\usepackage{nicefrac}       % compact symbols for 1/2, etc.
\usepackage{microtype}      % microtypography
\usepackage{lipsum}
\usepackage{graphicx}
\graphicspath{ {./images/} }

\title{Blockchain-based PKI within a Corporate Organization: Advantages and Challenges}

\author{
Julian Springer \\
  Department Computer Science and Security\\
  St. Pölten University of Applied Sciences\\
  St. Pölten, Austria \\
  \texttt{is221507@fhstp.ac.at} \\
   \And
 Philipp Haindl \\
  Department Computer Science and Security\\
  St. Pölten University of Applied Sciences\\
  St. Pölten, Austria \\
  \texttt{philipp.haindl@fhstp.ac.at} \\
}
\date{}
\usepackage[backend=biber,style=ieee, minbibnames=99,maxbibnames=99,mincitenames=1,maxcitenames=1, giveninits=false]{biblatex}

\begin{document}
\maketitle
\begin{abstract}
This research investigates the potential use of a blockchain-based Public Key Infrastructure (PKI) within an organization and compares it to conventional PKI systems. The goal is to assess the advantages and disadvantages of both approaches in order to determine the feasibility of employing blockchain technology for a decentralized PKI. The study will also evaluate the impact of current legal frameworks, such as the Cyber Resilience Act (CRA) and NIS-2 Directive. The study will examine various implementations of blockchain PKIs based on factors such as security, performance, and platform. The results indicate that blockchain-based PKIs can overcome the limitations of conventional PKIs by decentralizing the trust anchor, providing greater security. Blockchain technology allows for the immutable and transparent management of certificates, making tampering significantly more challenging. Additionally, blockchain-based PKIs offer enhanced mechanisms for identifying and addressing certificate misconduct.
\end{abstract}

\section{Objective}
It is assessed whether a public key infrastructure (PKI) based on blockchain technology offers an advantage within an association organization. Additionally, the challenges that arise are explained, and if feasible, how they can be addressed. This paper also includes the following contributions:
  \begin{itemize}
    \item  A fundamental examination of cryptography
    \item An introduction to the functionalities of both classic and blockchain-based PKIs
    \item The development of basic knowledge about how blockchains operate
    \item A list of current implementations of blockchain-based PKIs, along with their advantages and disadvantages in meeting the requirements
    \item A consideration of the impact of legal framework conditions
  \end{itemize}
  
  \section{Research Questions}
  The following research questions are answered in this paper:
  \begin{enumerate}
    \item What notable implementations are currently available and how effectively do they address the challenges posed by a traditional PKI?
    \item Is it practical to employ a decentralized PKI within a group organization?
    \begin{enumerate}
      \item Which distributed ledger technology would be appropriate for a decentralized Public Key Infrastructure (PKI)?
      \item How can one evaluate the suitability of employing a PKI?
    \end{enumerate}
    \item In what manner does the Cyber Resilience Act impact the execution of a blockchain-based Public Key Infrastructure (PKI)?
  \end{enumerate}

  \section{Cryptography Fundamentals}
  Cryptography is employed to ensure the confidentiality, integrity, authenticity, and non-repudiation of information. These four aspects guarantee that a third party cannot view the content of a message, the recipient can confirm that the message has not been altered during transmission, and the sender's identity can be verified. Non-repudiation, however, is not always guaranteed since there are protocols that do not offer this feature~\cite{hellwig_entwickeln_2021}.
  \subsection{Symmetric Cryptography}
  In current communication systems, only a single key is utilized by both parties involved in the exchange. However, transmitting this key securely poses a challenge. As the number of participants increases, the need for additional keys grows proportionately. The quantity of keys required can be determined by the formula $\frac{n*(n-1)}{2}$. One of the primary benefits of symmetric cryptography is its ability to secure substantial amounts of data with relatively short key lengths. Nevertheless, this approach cannot be applied to digital signatures because the key must be kept confidential at all times. In contemporary times, symmetric encryption is widely employed using the AES algorithm.
  \subsection{Asymmetric Cryptography}
  The primary distinction between symmetric cryptography and asymmetric cryptography lies in the utilization of a single shared key, which both participants must maintain in strict secrecy, and the employment of two keys per participant. These key pairs, also referred to as such, comprise of a private key and a public key, with the private key being consistently guarded in secrecy, while the public key is commonly acknowledged and frequently conveyed in the form of a certificate.
Unlike symmetric cryptography, there is no straightforward relationship between the encryption key, which is the public key, and the decryption key, which is the private key. This means that it is only possible to deduce the private key from the public key with significant effort. This phenomenon is rooted in mathematical functions that are easily computable but are practically impractical to compute without the knowledge of the private key. Examples of such mathematical functions, among others, include:
  \begin{itemize}
    \item Prime factorization
    \item Discrete logarithm
    \item Chinese remainder theorem
    \item Elliptical curves
  \end{itemize}
  Asymmetric cryptography presents several advantages, including the simplicity of distributing keys since the public key can generally be known and transmitted via insecure channels. Additionally, only one key pair is required per participant, reducing the number of such pairs. Asymmetric procedures also enable digital signatures using the private key instead of the public key, which is utilized in encryption. However, asymmetric cryptography has some drawbacks, such as less efficiency in the encryption process and the need for longer keys. Furthermore, it lacks security against quantum computers, unlike symmetric procedures. The most commonly used algorithms for asymmetric cryptography are RSA and elliptic curves.
  \subsection{Hybrid Cryptography}
  The customary approach incorporates both of the aforementioned techniques. Symmetrically, the message is encrypted. For the key exchange, the recipient discloses their public key, and the contents are encrypted asymmetrically. The recipient must first decrypt the symmetric key, which has been utilized to encrypt the content of the transmitted message, in order to access the content~\cite{komar_windows_2008}.
  \subsection{Digital signature}
The need for authenticity and integrity necessitates the implementation of specific measures. Such measures comprise:
\begin{itemize}
\item Technical security measures, such as digital signatures, which guarantee the authenticity of participants and the integrity of data by encrypting the hash value of the message with the sender's private key and attaching it to the message, which is then decrypted by the recipient using the sender's public key to ensure the authenticity and integrity of the data.
\item Certification authorities, which provide an infrastructure for secure key management.
\item Legal regulations, which ensure legal bindingness within an organization, although this is limited to the public sector and legally binding channels.
\end{itemize}

A digital signature functions as a substitute for a traditional signature, and there are functional, technical, and legal requirements that must be met in order to effectively implement digital signatures.
  \subsubsection{Functional requirements}
  Verifiability, forgery protection, and binding nature are essential aspects of a signature. These three elements are crucial to ensure that a signature is genuine and legally binding. Verifiability refers to the ability to verify the authenticity of the signature, while forgery protection ensures that only the owner of the signature can generate it. The binding nature of a signature means that the signatory cannot deny authorship once it has been signed. Therefore, it is critical to ensure that these elements are present in a signature to ensure its validity and legality.
  \subsubsection{Technical requirements}
The following points must be taken into consideration when evaluating the effectiveness of a digital signature solution:

\begin{itemize} 
  \item Performance: The verification process should not cause undue delay when loading the document.
\item  Security: The system must ensure cryptographic security to protect against unauthorized access or tampering.
\item  User Experience: The process of creating and verifying a signature should be designed to minimize inconvenience and facilitate ease of use.
\end{itemize}
  \subsubsection{Legal Requirements}
The legal effect should be equivalent to that of a traditional signature. It is crucial to ensure that the certification of the signatories is legally binding. Furthermore, the law must specify the requirements for the technical implementation and monitor compliance with them.
  
  \section{Introduction to Public Key Infrastructure}
  \subsection{Definition and Significance}
Public key infrastructure (PKI) comprises the various roles, policies, hardware, software, and procedures that are necessary to create, manage, distribute, and revoke digital certificates and manage public key encryption \cite{patil_evolving_2022}. The primary objective of PKI is to enable the secure electronic transmission of information. PKI allows users and computers to verify the legitimacy of the parties involved in electronic transactions, thus ensuring the confidentiality and integrity of the data exchanged. This is accomplished by issuing a digital certificate that can be used to identify a person or organization. In addition, directory services are utilized to store these certificates and to revoke them if necessary \cite{housley_public_2004, perlman_overview_1999}. As previously mentioned, key exchange remains a problem in larger environments. If the participants are aware of each other's public key, they can communicate via any public key-based protocol. Even if the public key does not need to be kept secret, indiscriminate distribution of it is not helpful, as it cannot be verified that a public key belongs to the entity it claims to be. For instance, a website could be hosted where public keys can be uploaded, and anyone can access it to obtain the necessary key. This would allow Bob to look up Alice's public key. However, Bob cannot be sure that it is actually Alice's key and not someone else's who wants to impersonate Alice. PKI enables the secure and convenient determination of public keys, which is essential for secure communication. It provides a mechanism for verifying the identities of the parties involved in electronic communication, facilitating trust within and between organizations and across the Internet \cite{perlman_overview_1999, patil_evolving_2022}.
  \subsection{Digital certificates}
  Digital certificates are formal documents that consist of a set of elements, as specified in the table, and are widely used to establish secure communication over the internet. These certificates provide assurance to the user that the public key contained within belongs to the specified subject, such as an individual, organization, or server~\cite{patil_evolving_2022}. The format and contents of digital certificates are determined by established standards. Table \ref{tab:x_509_digitales_zertifikat} lists the contents of a X.509 certificate.
  \begin{table*}[]
    \resizebox{\textwidth}{!}{%
      \begin{tabular}{@{}lll@{}}
        \toprule
        \textbf{Dataelement} & \textbf{X.509 Version} & \textbf{Explanation} \\ \midrule
        Version & 1 & An integer indicating the version of the certificate. \\ \midrule
        Serial number & 1 & \begin{tabular}[c]{@{}l@{}}An integer representing the unique number for each certificate issued by a certification authority.\end{tabular} \\ \midrule
        Signature & 1 & \begin{tabular}[c]{@{}l@{}}The identifier for the cryptographic algorithm used by the certification authority to sign the certificate.\\The value contains the identifier of the algorithm and all optional parameters that may be used by this algorithm.\end{tabular} \\ \midrule
        Issuer & 1 & The Distinguished Name (DN) of the certification authority that issued the certificate. \\ \midrule
        Validity period & 1 & The period (inclusive) during which the certificate is valid. \\ \midrule
        Subject & 1 & The Distinguished Name (DN) of the certificate applicant. \\ \midrule
        \begin{tabular}[c]{@{}l@{}}Information on the\\applicant's public key\end{tabular} & 1 & The public key held by the certificate applicant. \\ \midrule
        Unique ID of the issuer & 2 & \begin{tabular}[c]{@{}l@{}}A unique identifier representing the issuing certification body, as defined by the issuing certification body.\end{tabular} \\ \midrule
        Unique ID of the certificate applicant & 2 & \begin{tabular}[c]{@{}l@{}}A unique identifier that stands for the certificate applicant as defined by the issuing certification authority.\end{tabular} \\ \midrule
        Extentions & 3 & A collection of standard and Internet-specific certificate extensions. \\ \bottomrule
      \end{tabular}%
    }
    \caption{Data elements of a digital certificate according to the X.509 standard~\cite{kgremban_x509-zertifikate_2023}.}
    \label{tab:x_509_digitales_zertifikat}
  \end{table*}
  \section{Blockchain and Distributed Ledger}
  \subsection{Distributed Ledger}
  The primary objective of distributed ledger technologies (DLT) is to facilitate interactions between two parties without the need for a mutually trusted third party~\cite{el_ioini_review_2018}. DLTs such as blockchain offer a secure method for conducting and recording data transfers without a central administrative body. Distribution occurs through the shared use of copies of the ledger and their synchronization via a peer-to-peer network. New entries are added permanently in a cryptographically secure manner and made accessible to all participants~\cite{persons_blockchain_2019}. It is essential to recognize that a distributed ledger does not necessarily have to be implemented using a blockchain. A blockchain is simply one way of implementing a distributed ledger~\cite{rutland_blockchain_nodate}. Despite differences between various DLTs, such as the data model or technology, they all rely on three well-known technologies:
  \begin{itemize}
    \item Public key cryptography serves the purpose of providing secure identities for each participant, thereby enabling the ability to function in an insecure environment.
    \item A peer-to-peer distributed network serves the purpose of achieving scalability, avoiding single points of failure, and enabling decentralized control of the network by individuals.
    \item The consensus procedure is a means of achieving agreement among all participants on a common truth, without the need for a central authority that is universally trusted~\cite{el_ioini_review_2018}.
  \end{itemize}
  \subsection{Blockchain}
  As already pointed out, blockchain is a distributed ledger technology that involves the chronological chaining of blocks. When a participant wishes to conduct a transaction, they transmit all relevant information to all other network participants. A miner, a specialized node, then generates a block and attempts to calculate a valid hash value. Once a valid hash is found, all participants are informed and the validated block is added to the chain. This process is referred to as "mining" and is attractive to the financial sector, asset records, and identity management due to the immutability of validated data in the blockchain. A comparison of various distributed ledger technologies reveals that while all of these DLTs are theoretically possible, blockchain is the most suitable option due to its widespread use and resulting higher level of maturity, as well as its strong security features. The consensus mechanism's basic functioning is also well-suited to meet the requirements of a Public Key Infrastructure (PKI)~\cite{hellwig_entwickeln_2021}.
  
  \section{Decentralized PKI in a Group Organization}
  Several organizations are involved in a current PKI and have set themselves the goal of establishing a common trust anchor. The most important use case is private TLS certificates for interfaces between the individual organizations and their partners. With the help of common rules and standards, which are defined in the CP/CPS documents of the CAs, an improved quality of security and thus its increase should be achieved for the participating organizations.\\
  At the highest level, there is the root CA, which only issues certificates for sub-CAs. It is operated by an organization A, as agreed among themselves. Each of the participating organizations can operate one or more such sub-CAs. These are subject to the requirements of the policies of the Root CA and can tighten these as much as possible. Other participants (organizations) can use the services of the Root CA directly at any time by setting up a sub-CA. Partners can use the infrastructure indirectly by using certificates from the sub-CAs. It should also be noted that the participants can operate their own PKIs in parallel and independently. For reasons of controllability and separation of the technical areas, a separate root CA is created for each certificate type.
  \subsection{Criteria for evaluation of blockchain-based PKIs}
  The following factors can be utilized to assess the execution of a blockchain-based public key infrastructure (PKI):
  \begin{enumerate}
    \item Key feature: Each implementation possesses its own specialized focus, i.e. it targets a weakness of conventional PKI and proposes a solution.
    \item Blockchain type: The type of blockchain used, such as permissioned or permissionless.
    \item Blockchain platform: e.g. Ethereum, Bitcoin, or a custom implementation.
    \item Certificate: The format of the certificate, such as the X.509 standard or a custom format.
    \item Trust Model: The mechanism that selects the node responsible for validating a certificate, such as hierarchical or Web of Trust (WoT).
    \item Consensus mechanism: The type of consensus mechanism employed, such as proof-of-work or proof-stake.
    \item Storage: The method of storing blockchain data, which can be in the form of the entire data or the hash function of the block. There are two categories of blockchain data storage: on-chain storage, where the data is stored directly on the blockchain network, and off-chain storage, where the data is stored in a public ledger that can be accessed by all other nodes or in a private storage that can be accessed by the respective node.
    \item Time complexity: The algorithmic computational complexity in the form of time~\cite{panigrahi_smart_2022}.
  \end{enumerate}
  
  \section{Influence of CRA and NIS-2}
  The applicability of the CRA to the blockchain PKI hinges on the mode of delivery. If the PKI is provided as a service and customers are not required to install or host any software on their own, it is exempt from the CRA since SaaS solutions are covered under the NIS2 directive. Conversely, if the PKI software is sold as a product to be installed and operated by the purchaser, the CRA would apply~\cite{noauthor_cyber_2023}.
  
  \section{Conclusion}
  The study has revealed that blockchain-based Public Key Infrastructures (PKIs) possess the ability to tackle the fundamental flaws present in traditional PKIs. In particular, these include the reliance on central trust anchors, which can serve as a single point of failure in conventional PKIs. With the implementation of blockchain technology, the trustworthiness and authenticity of the certificates can be upheld in a decentralized manner, ensuring their immutability and enabling easy monitoring. This ultimately results in enhanced security, providing resistance against manipulation and potential failures.
  \subsection{Summary and Outlook}
  The research indicates that blockchain-based public key infrastructures (PKIs) present a promising alternative to traditional PKIs. They offer enhanced security, transparency, and resilience, which make them appealing for use in corporate organizations. However, implementing such a solution necessitates careful planning and evaluation to meet specific requirements and legal framework conditions.
  In the future, additional research and practical applications will be required to further optimize and improve the technology. Specifically, the feasibility of incorporating blockchain-based PKIs into existing IT infrastructures and the development of standards and best practices will be crucial to achieving widespread acceptance.
  Overall, this study makes a significant contribution to the discussion on the future of PKI and the role of blockchain technology in cybersecurity, beyond the financial and cryptocurrency sectors. It demonstrates that innovative approaches and technologies can effectively address the challenges of modern IT security.
  In conclusion, a blockchain-based PKI offers substantial advantages in terms of security and reliability, but selecting the appropriate distributed ledger technology and complying with legal requirements are crucial for the successful implementation of such a solution. These findings provide a solid foundation for further research and practical applications to fully leverage the benefits of blockchain technology in the field of PKI and enhance cybersecurity in corporate settings.
  
  \printbibliography

@book{komar_windows_2008,
	edition = {1},
	title = {Windows Server® 2008 {PKI} and Certificate Security},
	isbn = {978-0-7356-2516-7},
	pagetotal = {800},
	publisher = {Microsoft Press},
	author = {Komar, Brian},
	date = {2008-04-09},
}

@misc{rutland_blockchain_nodate,
	title = {Blockchain Byte},
	url = {https://www.finra.org/sites/default/files/2017_BC_Byte.pdf},
	publisher = {R3 Research},
	author = {Rutland, Emily},
	urldate = {2024-01-17},
}

@book{hellwig_entwickeln_2021,
	location = {Berlin, Heidelberg},
	title = {Entwickeln Sie Ihre eigene Blockchain: Ein praktischer Leitfaden zur Distributed-Ledger-Technologie},
	isbn = {978-3-662-62965-9},
	url = {https://link.springer.com/10.1007/978-3-662-62966-6},
	shorttitle = {Entwickeln Sie Ihre eigene Blockchain},
	publisher = {Springer Berlin Heidelberg},
	author = {Hellwig, Daniel and Karlic, Goran and Huchzermeier, Arnd},
	urldate = {2024-03-06},
	date = {2021},
	langid = {german},
	doi = {10.1007/978-3-662-62966-6},
}

@incollection{housley_public_2004,
	title = {Public Key Infrastructure ({PKI})},
	rights = {Copyright © 2004 John Wiley \& Sons, Inc. All rights reserved.},
	isbn = {978-0-471-48296-3},
	url = {https://onlinelibrary.wiley.com/doi/abs/10.1002/047148296X.tie149},
	booktitle = {The Internet Encyclopedia},
	publisher = {John Wiley \& Sons, Ltd},
	author = {Housley, Russ},
	urldate = {2024-04-04},
	date = {2004},
	langid = {english},
}

@article{perlman_overview_1999,
	title = {An overview of {PKI} trust models},
	volume = {13},
	issn = {1558-156X},
	url = {https://ieeexplore.ieee.org/document/806987},
	doi = {10.1109/65.806987},
	pages = {38--43},
	number = {6},
	journaltitle = {{IEEE} Network},
	author = {Perlman, R.},
	urldate = {2024-04-04},
	date = {1999-11},
	note = {Conference Name: {IEEE} Network},
}

@inproceedings{el_ioini_review_2018,
	location = {Cham},
	title = {A Review of Distributed Ledger Technologies},
	isbn = {978-3-030-02671-4},
	doi = {10.1007/978-3-030-02671-4_16},
	pages = {277--288},
	booktitle = {On the Move to Meaningful Internet Systems. {OTM} 2018 Conferences},
	publisher = {Springer International Publishing},
	author = {El Ioini, Nabil and Pahl, Claus},
	editor = {Panetto, Hervé and Debruyne, Christophe and Proper, Henderik A. and Ardagna, Claudio Agostino and Roman, Dumitru and Meersman, Robert},
	date = {2018},
	langid = {english},
}

@inproceedings{patil_evolving_2022,
	title = {Evolving Role of {PKI} in Facilitating Trust},
	url = {https://ieeexplore.ieee.org/document/9952249},
	doi = {10.1109/PKIA56009.2022.9952249},
	eventtitle = {2022 {IEEE} International Conference on Public Key Infrastructure and its Applications ({PKIA})},
	pages = {1--7},
	booktitle = {2022 {IEEE} International Conference on Public Key Infrastructure and its Applications ({PKIA})},
	author = {Patil, Vishwas T. and Shyamasundar, R.K.},
	urldate = {2024-04-11},
	date = {2022-09},
}

@online{kgremban_x509-zertifikate_2023,
	title = {X.509-Zertifikate},
	url = {https://learn.microsoft.com/de-de/azure/iot-hub/reference-x509-certificates},
	author = {kgremban},
	urldate = {2024-04-11},
	date = {2023-05-01},
	langid = {german},
}

@article{persons_blockchain_2019,
	title = {{BLOCKCHAIN} \& {DISTRIBUTED} {LEDGER} {TECHNOLOGIES}},
	url = {https://www.gao.gov/assets/gao-19-704sp.pdf},
	author = {Persons, Timothy M.},
	urldate = {2024-04-29},
	date = {2019-09},
	langid = {english},
}

@misc{panigrahi_smart_2022,
	title = {Smart Contract Assisted Blockchain based {PKI} System},
	url = {http://arxiv.org/abs/2207.09127},
	number = {{arXiv}:2207.09127},
	publisher = {{arXiv}},
	author = {Panigrahi, Amrutanshu and Nayak, Ajit Kumar and Paul, Rourab},
	urldate = {2024-05-21},
	date = {2022-07-19},
	langid = {english},
	eprinttype = {arxiv},
	eprint = {2207.09127 [cs]},
}

@online{noauthor_cyber_2023,
	title = {Cyber Resilience Act - Questions and Answers},
	url = {https://ec.europa.eu/commission/presscorner/api/files/document/print/en/qanda_22_5375/QANDA_22_5375_EN.pdf},
	urldate = {2024-05-21},
	date = {2023-12-01},
}

\end{document}